\documentclass[12pt]{iopart}

\usepackage{graphicx}
\begin{document}

\title{ALICE detector upgrades}

\author{Thomas Peitzmann for the ALICE Collaboration}

\address{Nikhef National Institute for Subatomic Physics, Amsterdam,
and Utrecht University,
P.O. Box 80000, 3508 TA Utrecht, The Netherlands.}
\ead{t.peitzmann@uu.nl}
\begin{abstract}

The LHC with its unprecedented energy offers unique opportunities for groundbreaking measurements in p+p, p+A and A+A collisions even beyond the baseline experimental designs. ALICE is setting up a program of detector upgrades, which could to a large extent be installed in the  LHC shutdown planned for 2017/18, to address the new scientific challenges. We will discuss examples of the scientific frontiers and will present the corresponding upgrade projects under study for the ALICE experiment.

\end{abstract}


\section{Introduction: the physics frontiers for ALICE}
Although ALICE has already demonstrated its excellent capabilities to measure high-energy nuclear collisions at the LHC, within its program to understand the properties of strongly interacting matter there are several frontiers for which the current experimental setup is not yet fully optimized. In fact, in some areas, where a lot of additional knowledge compared to that available during the design phase has been accumulated by previous experiments, detector upgrades will enhance the physics capabilities enormously. 
ALICE has in fact already upgraded its baseline apparatus by a transition radiation detector (TRD) and by electromagnetic calorimetry (EMCal), an enhancement of the latter for di-jet measurements (DCAL) has  recently been approved.
It is beyond the scope of this paper to comprehensively discuss the state-of-the-art of high-energy heavy-ion physics, so we will just mention a selection of important physics questions. 

Strong elliptic flow and jet quenching effects established at RHIC \cite{whitepapers} and already also observed at LHC \cite{lhc-hi} are arguably among the most important experimental results in heavy-ion physics of the last years. It has been seen that differential studies for different particle species are particularly powerful. The large baryon-to-meson ratio and the approximate quark number scaling in elliptic flow, which hint at the importance of quark degrees of freedom, call for particle identification capabilities. In addition, the description of jet quenching via parton energy loss will greatly profit from studies of in-medium fragmentation. Consequently, a full understanding of hadronization mechanisms, including coalescence and fragmentation, and possibly the role of chiral symmetry breaking, is crucial to understand particle production in these collisions and requires hadron identification in a large transverse momentum range.

Even more sensitive to details of medium modifications are particles carrying heavy flavour (charm and bottom). It is still unclear, how much heavy flavour shares the collective motion of the bulk matter. Also, there are indications of strong energy loss of heavy quarks in the medium, which can so far not be explained in models of QCD energy loss. A true understanding of these features, and also the production of heavy flavour will require identification of charmed hadrons to low transverse momenta and in particular of charmed baryons and multiple heavy hadrons. Furthermore, for the understanding of charmonium production a discrimination of prompt $J/\psi$ from those from $B$ decays is crucial.

Another interesting and as yet not fully understood phenomenon is the long-range rapidity correlation (aka "the ridge"), which was first observed in central Au+Au collisions at 200 $A$GeV by the STAR experiment, and has now been seen also at LHC by CMS 
\cite{ridge}. The long rapidity reach calls for a mechanism relevant in the earliest stages; at the same time the phenomenon seems to be related to the production of relatively high momentum particles ($\approx 6 \, \mathrm{GeV}/c$). Very different theoretical descriptions have been proposed, where the most promising approaches call for initial state density fluctuations (sometimes identified with flux tubes), which are visible in momentum space due to final state modifications (e.g. radial flow). However, there is not yet really a consistent theoretical description. 

At large rapidities, one also expects the influence of small-$x$ partons to become more and more important. This is the region where effects of gluon saturation \cite{saturation} should be most prominent. Signals consistent with gluon saturation have been observed at RHIC \cite{rhic-dau-raa}, but the interpretation is hampered by the very limited kinematical reach. The larger beam energy of the LHC will allow us to enter a new physics regime with access to much smaller values of $x$ and at the same time an enlarged phase space for saturation due to the expected larger saturation scale.

Thus further enhancements of particle identification will be crucial for the future physics program of ALICE. This applies both for hadrons made of light quarks, where measurements should extend out to higher $p_T$, and for hadrons containing heavy quarks, where coverage for baryons, multiply-heavy hadrons and in general towards lower $p_T$, as well as better control of $B$ feed-down is essential. Regarding measurements at large rapidity, or low $x$, ALICE will have a unique opportunity to enter a new physics regime, as the available space in the experimental setup allows for a flexible choice of the most adequate detector for forward measurements.

\section{Upgrade projects}
Heavy flavour measurements at central rapidity would be enhanced by implementing a new Inner Tracking System (ITS) with better secondary vertex resolution. The currently favoured design would consist of three layers of pixel and four layers of strip detectors using improved, thinner sensors with higher granularity. A crucial measure will be to place the innermost layer at a smaller distance from the beam ($R = 2.2 \,\mathrm{cm}$) while replacing the beam pipe with a smaller ($R = 2 \,\mathrm{cm}$) version. This would improve the secondary vertex resolution by a factor of $\approx 3$ and allow measurements down to much lower values of $p_T$. Furthermore the high efficiency and low contamination of the new ITS would allow for a level 2 trigger based on decay topology. For this upgrade two pixel technologies are being explored: hybrid pixels with a pixel size of $30 \, \mathrm{\mu m} \times 100 \, \mathrm{\mu m}$, and monolithic pixels with a pixel size of $20 \, \mathrm{\mu m} \times 20 \, \mathrm{\mu m}$. The sensors will have to tolerate a radiation dose of $> 2 \, \mathrm{Mrad}$ and should be read out in less than $50 \, \mathrm{\mu s}$. The detector should provide low level trigger information.

The largest limitation of the forward quarkonium measurements with the Muon Spectrometer is due to the inability to reconstruct possible secondary vertices of the muon tracks. In particular $B$-tagging of e.g. $J/\psi$ mesons would be crucial for the interpretation. This could be provided by additional tracking in front of the muon absorber, the Muon Forward Tracker (MFT), which would in addition improve mass resolution and background rejection. The MFT would likely use silicon tracking with technologies very similar to those used for the ITS upgrade, and also the integration of the two upgrades into the experiment would require strong cooperation of the two projects. One of the important design questions is whether the matching of MFT tracks with the tracks in the Muon Spectrometer is possible. First simulation studies indicate that this can be done very efficiently even in central Pb+Pb events.

Identification of light charged hadrons on a track-by-track basis could be performed with a new RICH detector. A suggested design builds on the experience obtained with the HMPID currently used in ALICE. The VHMPID\footnote{Very High Momentum Particle ID} detector would use a focusing RICH with spherical mirrors, a gaseous $C_4F_{10}$ radiator of 1~m length, a $CsI$-based photon detector and front-end electronics based on the Gassiplex chip. The detector should be able to identify hadrons well beyond $p_T = 10 \, \mathrm{GeV}/c$. The expected ranges for the identification of charged hadrons are given in Table~\ref{table1}. As an alternative option the usage of aerogel radiators is being investigated. To make efficient use of such a small acceptance detector a dedicated trigger detector may be implemented in addition.
\begin{table}[h]
\begin{center}
\begin{tabular}{|c|c|c|}
\hline 
$\pi$ & $K$ & $p$ \\ \hline
4 - 25 GeV/$c$ & 11 - 25 GeV/$c$ & 19 - 35 GeV/$c$ \\ \hline
\end{tabular}
\caption{Momentum ranges for particle identification with the VHMPID upgrade assuming an angular resolution of 1.2 mrad.}
\label{table1}
\end{center}
\end{table}

\begin{figure}[tbh]
\begin{center}
    \includegraphics[width=0.8\textwidth]{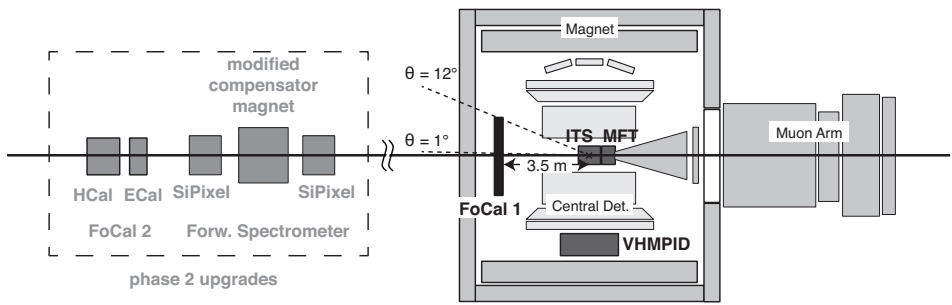}
\end{center}
    \caption{Schematic view of the ALICE setup. Indicated are locations of suggested detector upgrades as described in the text (ITS, MFT, VHMPID, and FoCal) as well possible later enhancements at more forward angles.}
    \label{figsetup}
\end{figure}

An obvious candidate for forward measurement with the generally higher momenta of particles in the lab system is an electromagnetic calorimeter with its intrinsically better resolution at higher energies and the possibility to identify photons and neutral hadrons. A possible location of such a detector would be on the A-side (opposite the Muon Spectrometer, see Figure~\ref{figsetup}) at a distance of $3.5 \, \mathrm{m}$, replacing the currently installed Photon Multiplicity Detector. While the demands on energy resolution may be moderate at this location, the small opening angle of neutral pion decays and the overall large particle density in Pb+Pb collisions will require good position resolution and two-particle separation power. To achieve this, the Moli\`ere radius of the material should be small and the granularity of the signal readout high. The favourite design for the Forward Calorimeter (FoCal) would use $W$ ($R_M = 9 \, \mathrm{mm}$) as absorber and Si-sensors as active material, which would make the desired granularity of $\approx 1\times 1 \, \mathrm{mm}^2$ possible. As sensor technologies, conventional pad/pixel sensors with separate readout and monolithic pixels will be investigated.

These upgrades (phase 1) would very likely be installed in the long LHC shutdown period in 2017/2018. Further more sophisticated upgrades (phase 2) will be studied, e.g. a comprehensive forward detector at still larger pseudorapidities ($ 5 < \eta < 7$) consisting of electromagnetic and hadronic calorimetry and charged particle tracking. Such a detector, which would be located at a larger distance from the vertex (see Figure~\ref{figsetup}), would need a major redesign of the setup and beam infrastructure and would more likely be geared towards a later shutdown period.

\end{document}